\documentclass[journal,twocolumn,10pt,twoside]{IEEEtran}

\ifCLASSINFOpdf
\else
\fi

\usepackage{multirow}
\usepackage{booktabs}
\usepackage{amssymb}
\usepackage{graphicx}
\usepackage[cmex10]{amsmath}
\usepackage{array}
\usepackage{mdwtab}
\usepackage{fixltx2e}
\usepackage{graphics}
\usepackage{epsfig}
\usepackage{mathptmx}
\usepackage{times}
\usepackage{amsmath}
\usepackage{amssymb}
\usepackage{url}
\usepackage{lscape}
\usepackage{color}
\usepackage{epsfig}

\usepackage{cite}
\usepackage{balance}
\usepackage{dblfloatfix}
\usepackage{nicefrac}
\usepackage{hyperref}

\begin{document}

\title{Constrained Control of Depth of Hypnosis During Induction Phase}

\author{Mehdi~Hosseinzadeh,~\IEEEmembership{Member,~IEEE,}~Guy~A.~Dumont,~\IEEEmembership{Life~Fellow,~IEEE,}~and~Emanuele~Garone,~\IEEEmembership{Member,~IEEE}
\thanks{This research has been funded by the FNRS MIS ``Optimization-free Control of Nonlinear Systems subject to Constraints", Ref. F.4526.17.}
\thanks{M. Hosseinzadeh and E. Garone are with the Service d'Automatique et d'Analyse des Syst\`{e}mes  (SAAS), Universit\'{e} Libre de Bruxelles (ULB), Brussels, Belgium (email: mehdi.hosseinzadeh@ieee.org; egarone@ulb.ac.be).}
\thanks{G. A. Dumont is with the Department of Electrical and Computer Engineering, The University of British Columbia, Vancouver, Canada (email: guyd@ece.ubc.ca).}
}
\maketitle

\begin{abstract}
This paper proposes a constrained control scheme for the control of the depth of hypnosis during induction phase in clinical anesthesia. In contrast with existing control schemes for propofol delivery, the proposed scheme guarantees overdosing prevention while ensuring good performance. The core idea is to reformulate overdosing prevention as a constraint, and then use the recently introduced Explicit Reference Governor to enforce the constraint satisfaction at all times. The proposed scheme is evaluated in comparison with a robust PID controller on a simulated surgical procedure for 44 patients whose Pharmacokinetic-Pharmacodynamic models have been identified using clinical data. The results demonstrate that the proposed constrained control scheme can deliver propofol to yield good  induction phase response while preventing overdosing in patients; whereas other existing schemes might cause overdosing in some patients. Simulations show that mean rise time, mean settling time, and mean overshoot of less than 5 [min], 8 [min], and 10\%, respectively, are achieved, which meet typical anesthesiologists' response specifications.
\end{abstract}

\section{Introduction}\label{Introduction}
\IEEEPARstart{A}{NESTHESIA} means lack of ability to sense, or a state of being unable to feel anything. More precisely, it is a temporary reversible state consisting of unconsciousness, loss of recall, lack of pain perception, and sometimes muscle relaxation. During surgical procedures, anesthesiologists adjust the dose of administered anesthetic drug to reach an acceptable level of anesthesia. From technical viewpoint, their actions can be interpreted as a manual feedback control. Starting from this observation in recent years automating drug delivery in anesthesia has gained more and more  attention.

Automated drug delivery in anesthesia involves the continuous administration of an anesthetic drug to achieve loss of consciousness while maintaining safe vital signs during the surgery. In general, anesthesia consists of three components \cite{Bibian2005}: (i) hypnosis (\textit{i.e.,} loss of consciousness  and lack of awareness), (ii) analgesia (\textit{i.e.,} lack of nociceptive reactivity), and (iii) neuromuscular blockade (\textit{i.e.,} hemodynamic stability and immobilization). This paper will deal with the control of depth of hypnosis through intravenous administration of propofol. In other words, the main goal of this paper is to propose a delivery system that can safely control the depth of hypnosis by manipulating the infusion rate of propofol.

Propofol hypnosis can be divided into three temporal phases \cite{Soltesz2011}: (i) induction, (ii) maintenance, and (iii) emergence. The aim of the induction phase is to bring the patient from total awareness to a desired depth of hypnosis. Once a stable depth of hypnosis is achieved, the maintenance phase begins. Surgery takes place during the maintenance phase. After completing the surgery, the emergence phase begins, when the administration of propofol is terminated.

Although controlling the depth of hypnosis during the maintenance phase is onerous \textit{per se}, one of the main challenges in propofol delivery is to safely administer the drug during the induction phase despite the patients' inherent drug response variability without overdosing them. In this paper, we will focus on the development of a constrained control scheme to guarantee overdosing prevention while ensuring an acceptable hypnosis induction performance.

In recent years a large number of controllers for the control of the depth of hypnosis have been proposed in the literature. Indeed, the recent development of devices and techniques to quantify the depth of hypnosis \textit{e.g.,} the bispectral index \cite{Liu1997} and wavelet-based index \cite{Zikov2006_2}, has opened up the opportunity for developing automatic propofol delivery systems. Several control schemes have been proposed for the control of the closed-loop control of hypnosis including  PID controller \cite{Absalom2002,Dumont2009,Heusden2014,Padula2017_1}, $H_\infty$ controller \cite{Hahn2012,Caiado2013,Lemos2014}, multi-model robust control scheme \cite{Hosseinzadeh2018}, and event-based controller \cite{Padula2017_2}

The most important confounders in closed-loop hypnosis control are \cite{Hahn2012}: (i) the large amount of inter-individual model variability, and (ii) the persistent and unexpected surgical stimulation and anesthetic-analgesic interaction throughout the surgical procedure. The former is an onerous problem during both induction and maintenance phases, while the latter is an aspect that is specific to the maintenance phase. It is obvious that, because of these two confounders, a purely model-based control scheme cannot effectively control the depth of hypnosis or may even cause instability. Hence robustness against those confounders is the main matter of concern in control schemes presented to control the depth of hypnosis.

Although fundamental for the safety of the patient, to the best of authors' knowledge, no scheme to systematically ensure overdosing prevention during the induction phase has been proposed yet. Indeed, existing works state that the risk of overdosing decreases by increasing the robustness of the system \cite{Hahn2012,Hosseinzadeh2018}, or by adding a set-point prefilter to smooth the reference signal and reduce possible overshoots \cite{Dumont2009}. At the current stage the absence of schemes guaranteeing overdosing prevention is one of the aspects that limits the acceptance of these schemes in the anesthesiology community.

In this paper we will first reformulate the control of depth of hypnosis during induction phase as a constrained control problem. Then we will propose a control architecture to control the induction phase guaranteeing overdosing prevention. In particular, we will make use of the recently introduced Explicit Reference Governor (ERG) framework \cite{Nicotra2015_1,Hosseinzadeh2018ERG,Garone2016_2,Garone2017,Nicotra2018}. The main idea behind the ERG framework is to determine an invariant set that would contain the state trajectory if the currently auxiliary reference were to remain constant. If the distance between this invariant set and the boundary of the constraints is strictly positive, it follows from continuity that the derivative of the auxiliary reference can be nonzero without leading to constraint violations. If this distance is zero, the satisfaction of the constraints is ensured by maintaining the current reference constant. One of the main strengths of the ERG is that it requires very limited computational capabilities since, unlike other constrained control schemes (\textit{e.g.,} Model Predictive Control), it does not make use of online optimization, making its implementation simple, robust, and easily certifiable.

The rest of the paper is organized as follows. Section \ref{modeling} describes the models used in propofol delivery system. Section~\ref{control_architecture} presents the details of the proposed control scheme. In Section~\ref{simulation},  simulations are carried out using the proposed scheme and their results are discussed. Finally, Section~\ref{conclusion} concludes the paper.
\section{Modeling of Propofol Delivery System}\label{modeling}
The relationship between dose and pharmacological effect of administered anesthetic drug (propofol in this paper) is described by the PharmacoKinetic-PharmacoDynamic (PKPD) model. The input to this process is the dose of administered drug, and the output is the clinical hypnotic effect. For the sake of completeness and for the reader's convenience, the overall PKPD model will be hereafter described.

\subsection{PKPD Model}
The model which is normally used to explain the response of a patient to administered aesthetic drug (propofol in this study) consists of two parts: (i) PK model, and (ii) PD model. The PK model relates the drug plasma concentration with the administered dose. Most propofol PK models consider three compartments \cite{Schuttler}: (i) plasma compartment, (ii) shallow peripheral compartment, and (iii) deep peripheral compartment. Denoting the propofol concentration in the plasma, fast peripheral, and slow peripheral compartments as $C_1$, $C_2$, and $C_3$ (all in [mg/l] or [$\mu$g/ml]), respectively, and the volume in the aforementioned compartments as $V_1$, $V_2$, and $V_3$ (all in [l]), respectively, the state-space representation of the PK model can be expressed as
\begin{align}\label{eq:PKmodel}
\left[\begin{matrix}\dot{C}_1 \\ \dot{C}_2 \\ \dot{C}_3\end{matrix}\right]=&\left[\begin{matrix} -(k_{10}+k_{12}+k_{13}) & k_{12} & k_{13} \\ k_{21} & -k_{21} & 0 \\ k_{31} & 0 & -k_{31} \end{matrix}\right]\left[\begin{matrix}C_1 \\ C_2 \\ C_3\end{matrix}\right]+\left[\begin{matrix}\frac{1}{V_1} \\ 0 \\ 0 \end{matrix}\right]I,
\end{align}
where where $I(t)$ is the infusion rate (in [mg/s]), and
\begin{equation}
k_{10}=\frac{Cl_1}{V_1},~k_{12}=\frac{Cl_2}{V_1},~k_{21}=\frac{Cl_2}{V_2},~k_{13}=\frac{CL_3}{V_1},~k_{31}=\frac{Cl_3}{V_3},
\end{equation}
and $Cl_1$ denotes the elimination clearance, and $Cl_2$ and $Cl_3$ are inter-compartmental clearances. Note that $V_i$ and $Cl_i,~i=1,2,3$ are determined using the relations presented in \cite{Schuttler}.

\begin{figure}[!t]
  \centering
  \includegraphics[width=7cm]{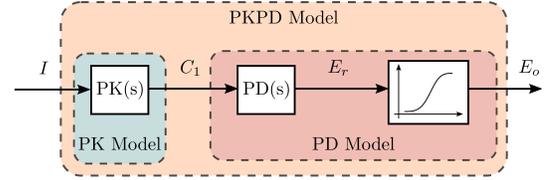}
  \caption{PKPD model block diagram.}\label{PKPDmodel}
\end{figure}

The PD model relates the plasma concentration with the pharmacological end-effect. In other words, the PD model describes the distribution of the propofol in the brain. The PD model is a first-order plus time-delay system \cite{BibianPhDThesis}, as
\begin{equation}\label{eq:PDmodel}
PD(s)=\frac{E_r(s)}{C_1(s)}=e^{-T_ds}\frac{1}{2EC_{50}}\frac{k_d}{s+k_d},
\end{equation}
where $T_d$ and $k_d$ are the transport delay and rate of propofol distribution between the plasma concentration and the brain, and $EC_{50}$ is the steady-state plasma concentration to obtain 50\% of the hypnotic effect. In addition, a nonlinear saturation function (Hill function) is used to describe the relation between $E_r(t)$ and the clinical hypnotic effect $E_o(t)$, which is
\begin{equation}\label{eq:Hill}
E_o(t)=\frac{\left(E_r(t)\right)^\gamma}{0.5^\gamma+\left(E_r(t)\right)^\gamma},
\end{equation}
where $\gamma$ is the cooperativity coefficient (see \cite{Ionescu2018} for details on how to simplify the number of parameters in the Hill curve). Note that $E_o(t)$ is bounded between 0 and 1; 0 means no hypnotic effect, 1 is associated with maximum hypnosis.

Combining \eqref{eq:PKmodel} and \eqref{eq:PDmodel}, and linearizing \eqref{eq:Hill} in the neighborhood of the operating regime, \textit{i.e.}, $E_r(t)=0.5$ (see Appendix), the following PKPD model is obtained, which describes the drug dose-hypnotic effect relationship of propofol:
\begin{equation}\label{PKPD_model}
\text{PKPD}(s)=e^{-T_ds}K_{PKPD}\frac{(s+z_1)(s+z_2)}{(s+p_1)(s+p_2)(s+p_3)(s+p_4)},
\end{equation}
where $K_{PKPD}$, $z_i,~i=1,2$, and $p_j,~j=1,\cdots,4$ are the gain, zeros, and poles of the model, respectively.

\subsection{$\text{WAV}_{\text{CNS}}$ Monitor Model}
In this paper, we assume that the clinical hypnotic effect is measured through the $\text{WAV}_{\text{CNS}}$ \cite{Zikov2006}. The dynamics of the $\text{WAV}_{\text{CNS}}$ monitor is usually modeled as \cite{Dumont2009}
\begin{equation}\label{eq:sensor}
H(s)=\frac{Y(s)}{E_o(s)}=\frac{1}{(8s+1)^2},
\end{equation}
where $Y(s)$ is the Laplace transform of the $\text{WAV}_{\text{CNS}}$ index.

\subsection{Age Groups}
In this paper, we will consider the  PK and PD model data identified for 44 patients in \cite{Dumont2009}. The data consists of two data sets: (i) the patient characteristics including gender, age, weight, and height, which are used to calculate the PK parameters, and (ii) the estimated PD parameters including $T_d$, $k_d$, $EC_{50}$, and $\gamma$, which are used in the PD model directly as in \eqref{eq:PDmodel}. As discussed in Section~\ref{Introduction}, the main difficulty in designing an automatic drug delivery system for anesthesia is the inherent patient variability. As discussed in \cite{Schnider1999}, patient age can be used as a criterion to reduce the inter-individual variability of the PKPD models. Hence, a 10-years bracket is selected and the 44 patients are subdivided into four age groups, as Group 1: 18-29 years, Group 2: 30-39 years, Group 3: 40-49 years, and Group 4: 50-60 years.

A nominal model for each age group is identified using the optimization procedure presented in \cite{Dumont2009}. The obtained nominal parameters are presented in \tablename~\ref{Table:Nominal_parameters}.

\section{Control Architecture}\label{control_architecture}
This section discusses the development of a constrained control scheme for the control of depth of hypnosis. Following the usual Reference Governor philosophy \cite{Garone2016_1}, the procedure consists in first pre-stabilizing the system; then augmenting it with an \textit{add-on} unit to enforce constraints satisfaction.

\subsection{Pre-Stabilizing the Propofol delivery system}
\begin{figure}[!t]
  \centering
  \includegraphics[width=8cm]{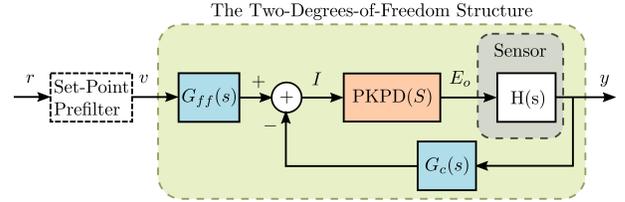}
  \caption{The two-degrees-of-freedom control scheme presented in \cite{Dumont2009}.}\label{Fig:Structure_PID}
\end{figure}

To stabilize the propofol delivery system, the robust PID controller designed in \cite{Dumont2009} will be used. The controller is implemented with two degrees of freedom as depicted in \figurename~\ref{Fig:Structure_PID}, where $G_c(s)$ and $G_{ff}(s)$ are expressed as:
\begin{align}
G_{ff}(s)=&k_p+\frac{k_i}{s},\label{eq:Gff}\\
G_c(s)=&k_p+\frac{k_i}{s}+k_ds.\label{eq:Gc}
\end{align}

The numerical values of the parameters for each group are reported in \tablename~\ref{Table:PID_parameters}. Note that since it is necessary to protect the controller from integrator windup, particularly when the infusion rate is nil, a back-calculation anti-windup scheme is implemented that resets the integrator dynamically with a time constant $T_t$. It should be remarked that the controller parameters are determined based on nominal parameters presented in \tablename~\ref{Table:Nominal_parameters} for each age group. As shown in \cite{Soltesz2011,Keyser2015}, patient-individualized schemes can improve the performance of the system, which is out of the scope of this paper.

For notational compactness, the overall dynamic model of the pre-stabilized system (green box in \figurename~\ref{Fig:Structure_PID}) will be denoted as
\begin{equation}\label{eq:model_nonlinear}
\left\{
\begin{array}{l}
\dot{x}(t)=f(x(t),v(t))\\
y(t)=h(x(t),v(t))
\end{array}
\right.,
\end{equation}
where $x\in\mathbb{R}^{n}$ is the state of the pre-stabilized system, $y(t)\in[0,1]$ is the output of the system representing the current level of hypnosis, and $v\in[0,1]$ is the desired reference of the control loop, \textit{i.e.,} the desired level of hypnosis. Note that in order to build \eqref{eq:model_nonlinear}, the actual PKPD model given in \eqref{eq:PKmodel}-\eqref{eq:Hill} is used.

\begin{table}[!t]
\centering
\caption{Optimum Nominal $\text{PKPD}(s)$ Parameters.}\label{Table:Nominal_parameters}
\footnotesize
\begin{tabular}{c|c|c|c|c}
Parameters & Group 1 & Group 2 & Group 3 & Group 4 \\
\hline
$T_d$ [s] & 18.6 & 16.5 & 8.3 & 17.8\\
$K_{PKPD}$ ($10^{-4}$) [\nicefrac{1}{mg$\text{s}^{-1}$}] & 1.698 & 1.928 & 1.438 & 1.823 \\
$z_1$ ($10^{-3}$) [s$^{-1}$] & 1.477 & 1.478 & 1.486 & 1.478 \\
$z_2$ ($10^{-5}$) [s$^{-1}$] & 2.572 & 2.703 & 3.627 & 2.651 \\
$p_1$ ($10^{-2}$) [s$^{-1}$] & 3.239 & 3.843 & 2.870 & 3.656 \\
$p_2$ ($10^{-3}$) [s$^{-1}$] & 6.961 & 7.735 & 7.748 & 9.100 \\
$p_3$ ($10^{-4}$) [s$^{-1}$] & 2.803 & 2.912 & 2.843 & 2.962 \\
$p_4$ ($10^{-5}$) [s$^{-1}$] & 2.703 & 2.787 & 2.121 & 2.710
\end{tabular}
\end{table}

\begin{table}[!t]
\centering
\caption{PID Parameters.}\label{Table:PID_parameters}
\footnotesize
\begin{tabular}{c|c|c|c|c}
Parameters & Group 1 & Group 2 & Group 3 & Group 4 \\
\hline
$k_p$ & 2.610 & 3.947 & 10.207 & 4.455 \\
$k_i$ & 0.026 & 0.046 & 0.107 & 0.058 \\
$k_d$ & 65.09 & 85.29 & 202.38 & 104.83 \\
$T_t$ [s] & 49.819 & 43.024 & 43.397 & 42.416 \\
\end{tabular}
\end{table}

\subsection{Enforcing Constraint Satisfaction}
The two-degrees-of-freedom structure presented in the previous sub-section is sufficiently robust against intra- and inter-patient variabilities, and guarantees reference tracking in a reasonable time. However it may cause high overshoot in the induction phase, which increases the risk of overdosing. Note that for safety $y(t)$ should be less than 0.6 for all times.

In \cite{Dumont2009}, to cope with this problem, a low-pass set-point prefilter was introduced
\begin{equation}\label{eq:passive_prefilter}
F_{sp}(s)=\frac{V(s)}{R(s)}=\frac{1}{T_{sp}s+1},
\end{equation}
where $T_{sp}$ is the time constant of the filter whose values are reported in \tablename~\ref{Table:Prefilter_parameter}. The idea is that the set-point prefilter \eqref{eq:passive_prefilter} generates an auxiliary reference
$v(t)$ by smoothing (reducing the sharpness of) the reference $r(t)$. This decreases the overshoot, and consequently reduces the overdosing occurrences. However, it does not guarantee overdosing prevention. To guarantee overdosing prevention in all cases one should increase the time constant $T_{sp}$ that will lead to a very slow response. In the next subsection, an analytic approach to design an active set-point prefilter to guarantee overdosing prevention at all times without hampering the performance in an unacceptable way will be presented.

\begin{table}[!t]
\centering
\caption{Passive Set-Point Prefilter Parameter.}\label{Table:Prefilter_parameter}
\footnotesize
\begin{tabular}{c|c|c|c|c}
Parameter & Group 1 & Group 2 & Group 3 & Group 4 \\
\hline
$T_{sp}$ [s] & 156.81 & 129.90 & 111.95 & 124.96 \\
\end{tabular}
\end{table}

\subsection{ERG-Based Active Set-Point Prefilter}
The ERG framework \cite{Garone2017} is an an \emph{add-on} unit which acts as an active set-point prefilter. The ERG suitably modifies the derivative of the auxiliary reference $v(t)$, only when it is needed, such that constraints are enforced at all times. This implies that using the ERG scheme as the set-point prefilter in the structure presented in \figurename~\ref{Fig:Structure_PID} can guarantee not only overdosing prevention but also acceptable performance during the induction phase. In this subsection we will propose a ERG scheme specifically designed for the control of the depth of hypnosis.

Once the propofol delivery system is stabilized by the proposed two-degrees-of-freedom structure \cite{Dumont2009}, the next step is to add the constraint-handling capability to make sure that $y(t)\leq0.6$ for all times. As shown in \cite{Nicotra2018}, this can be done by manipulating the auxiliary reference $v(t)$ according to the following differential equation
\begin{equation}\label{ClassicERG}
\dot{v}(t)=\kappa\cdot\Delta(t)\cdot\rho(t),
\end{equation}
where $\kappa>0$ is a tuning parameter, and $\Delta(t)$ and $\rho(t)$ are the two fundamental components of the ERG scheme, called the Dynamic Safety Margin (DSM) and the Navigation Field (NF), respectively.

The NF represents the direction along a feasible path that leads from the current auxiliary reference hypnosis level $v$ to the desired hypnosis level $r$. In other words, the NF can be interpreted as the answer to the question ``\textit{What direction should the auxiliary reference hypnosis level follow?}" \cite{Nicotra2018}. In mathematical terms, $\rho: \mathbb{R}\times\mathbb{R}\rightarrow\mathbb{R}$ is a NF if, for any strictly admissible constant desired hypnosis level $r\in\mathbb{R}$ and for any admissible initial value $v(t_0)$ such that $h(\overline{x}_{v(t_0)},v(t_0))\leq0.6$ where $\overline{x}_{v(t_0)}$ denotes the equilibrium of \eqref{eq:model_nonlinear} associated to $v(t_0)$ (\textit{i.e.}, $f(\overline{x}_{v(t_0)},v(t_0))=0$), the system $\dot{v}(t)=\rho(t)$ satisfies
\begin{align}
&h(\overline{x}_{v(t)},v(t))\leq0.6,~\forall t\geq t_0,\\
&\lim\limits_{t\rightarrow\infty}v(t)=r.
\end{align}

Since in closed-loop anesthesia the reference is mono-dimensional, it is sufficient to choose the NF as
\begin{equation}\label{rho}
\rho(t)=\frac{r-v(t)}{\max\{|r-v(t)|,\eta\}},
\end{equation}
where $\eta>0$ is a smoothing factor.

The DSM represents a distance between the constraint $y(t)\leq0.6$ and the output of the system \eqref{eq:model_nonlinear} that would emanate from the state $x(t)$ for a constant reference hypnosis level $v$. In other words, the DSM can be interpreted as the answer to the question ``\textit{How safe is it to change the auxiliary reference hypnosis level $v$?}" \cite{Nicotra2018}. In mathematical terms, $\Delta: \mathbb{R}\times\mathbb{R}^n\rightarrow\mathbb{R}$ is a DSM if, for a given auxiliary reference hypnosis level $v$ satisfying $h(\overline{x}_v,v)\leq0.6$, the following properties hold true:
\begin{itemize}
\item $\Delta(t)$ is continuous and bounded for any bounded $\|x\|$ and $|v|$.
\item $\Delta(t)\geq0$ implies that if the current auxiliary reference hypnosis level $v$ is maintained constant, $y(t)=h(x(t),v)\leq0.6$ for all times.
\item $\Delta(t)>0$ implies that $v(t)$ can be perturbed without causing constraint violation.
\end{itemize}

The mentioned properties mean that for given a constant reference $v$, $\Delta(t)$ is a \textit{distance} representing how far $y(t)$ is from 0.6.

As shown in \cite{Nicotra_phd,Nicotra2018}, the most intuitive way to compute the DSM is solving at each time instant the following initial value problem
\begin{equation}\label{eq:prediction}
\left\{
\begin{array}{l}
\dot{\hat{x}}(\tau)=f(\hat{x}(\tau),v(t))\\
\hat{x}(0)=x(t)
\end{array}
\right.,
\end{equation}
and then assigning $\Delta(t)$ as
\begin{equation}\label{Delta1}
\Delta(t)=\min_{\tau\in\left[t,\infty\right)}\{0.6-\hat{y}(\tau|t)\},
\end{equation}
where $\hat{y}(\tau|t)$ is
\begin{equation}\label{output_nonlinear}
\hat{y}(\tau|t)=h(\hat{x}(\tau|t),v(t)).
\end{equation}


Clearly, solving \eqref{eq:prediction} at each time instant over an infinite horizon is inapplicable in practice. To make the approach applicable, the DSM \eqref{Delta1} can be reformulated as
\begin{equation}\label{Delta1}
\Delta(t)=\min_{\tau\in\left[t,t+T\right]}\{0.6-\hat{y}(\tau|t)\},
\end{equation}
where $T$ is a finite time instant such that $\hat{x}(t+T|t)\in\Omega$, in which $\Omega$ is an invariant set included in $\|\hat{x}(\tau)-\overline{x}_{v(t)}\|\leq\varepsilon$, with $\varepsilon>0$ as a design parameter and $\overline{x}_{v(t)}$ as the equilibrium point of system \eqref{eq:prediction} (\textit{i.e.,} $f(\overline{x}_{v(t)},v(t))=0$).
For more details on determining the set $\Omega$ and calculating the time instant $T$, please refer to \cite{Nicotra_phd,Nicotra2018}.

The aforementioned method to restrict simulation of system \eqref{eq:prediction} to an finite horizon $T\geq0$ does not compromise the properties of the DSM \cite{Nicotra_phd} and can be used in practice easily. However, since $\hat{x}(t+T|t)\in\Omega$ is the stopping condition for the trajectory prediction, it is concluded that the prediction horizon $T$ is not constant. This means that the execution time of \eqref{Delta1} is not known \textit{a priori}, which can be problematic in real-time applications.

Observing results of extensive simulations with different initial conditions and hypnosis reference levels demonstrates that \eqref{eq:model_nonlinear} can be well approximated with a linear system. Clearly, for linear systems any constant value bigger than the peak time (the time at which peak value occurs) of the system can be chosen as the prediction horizon. Also, computing $\hat{y}(\tau|t),~\tau\in[t,t+T]$ for a linear system with initial condition $x(t)$ and input $v(t)$ is easier and less computationally intensive than \eqref{output_nonlinear}. Therefore,  in the following, first the system \eqref{eq:model_nonlinear} will be approximated by a linear system; then, a procedure will be presented to modify the resulting DSM in order to guarantee constraints satisfaction at all times in the presence of approximation error.

From extensive simulations, it is concluded that the system \eqref{eq:model_nonlinear} can be well approximated with the following linear system
\begin{equation}\label{eq:linear_model}
\left\{
\begin{array}{l}
\dot{\tilde{x}}(t)=A\tilde{x}(t)+Bv(t)\\
\tilde{y}(t)=C\tilde{x}(t)+Dv(t)
\end{array}
\right.
\end{equation}
where
\begin{align}
A=&\left[\begin{matrix}A_1 & 0 & 0 & 0\\0 & A_2 & 0 & B_2C_4\\ B_3C_1 & -B_3C_2 & A_3 & -B_3D_2C_4 \\ 0 & 0 & B_4C_3 & A_4\end{matrix}\right],\label{Alinear}\\
B=&\left[\begin{matrix}B_1^T & 0 & \left(B_3D_1\right)^T & 0 \end{matrix}\right]^T,\label{Blinear}\\
C=&\left[\begin{matrix}0 & 0 & 0 & C_4\end{matrix}\right],\label{Clinear}\\
D=&0,
\end{align}
with $(A_1,B_1,C_1,D_1)$, $(A_2,B_2,C_2,D_2)$, $(A_3,B_3,C_3,D_3)$, and $(A_4,B_4,C_4,D_4)$ as state-space realization matrices of the feedforward controller $G_{ff}(s)$ given in \eqref{eq:Gff}, the feedback controller $G_{c}(s)$ given in \eqref{eq:Gc}, the PKPD model \eqref{PKPD_model}, and  the sensor \eqref{eq:sensor}, respectively. Note that to compute $(A_3,B_3,C_3,D_3)$, the time-delay operator in \eqref{PKPD_model} is approximated by Pad\`{e} approximant. Therefore, $\tilde{y}(\tau|t)$ can be computed as
\begin{equation}\label{xbar_closedsolution}
\tilde{y}(\tau|t)=Ce^{A(\tau-t)}x(t)+C\int_t^\tau \left(e^{A(\tau-\sigma)}Bv(t)\right)d\sigma,
\end{equation}
for $\tau\in[t,t+T]$, where $T$ is bigger than the peak time of system \eqref{eq:linear_model} for all possible initial conditions (see \cite{Garone2016_1} for details on how to compute this peak time). In this study, it is assumed that $T=300~[s]$, which is sufficiently larger than the peak time of system \eqref{eq:linear_model} computed for the 44 patients.

At this point, define the approximation error as $\hat{e}(t)\triangleq y(t)-\tilde{y}(t)$. Thus, we can define $\delta_0$ as a measure of approximation accuracy as
\begin{equation}
\delta_0\triangleq\max\limits_{G_i}\,\sup\limits_t|\hat{e}(t)|,
\end{equation}
where $G_i,~i=1,\cdots,4$ is the set of all patients of the \textit{i}-th group. Thus, in order to guarantee constraint satisfaction in the presence of approximation error, we could restrict the DSM \eqref{Delta1} as follows
\begin{equation}\label{Delta1_1}
\Delta(t)=\min_{\tau\in\left[t,t+T\right]}\{0.6-\tilde{y}(\tau|t)-\delta_0\}.
\end{equation}
where $\delta_0$ can be interpreted as safety bound to take into account the mismatch between \eqref{eq:prediction} and \eqref{eq:linear_model}.

\begin{figure}[!t]
  \centering
\includegraphics[width=4.5cm]{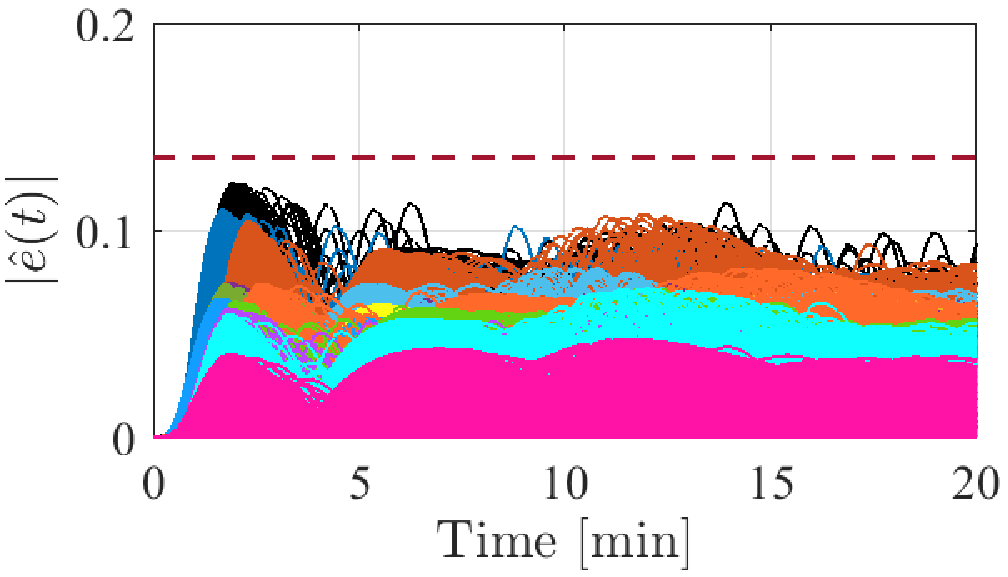}\includegraphics[width=4.5cm]{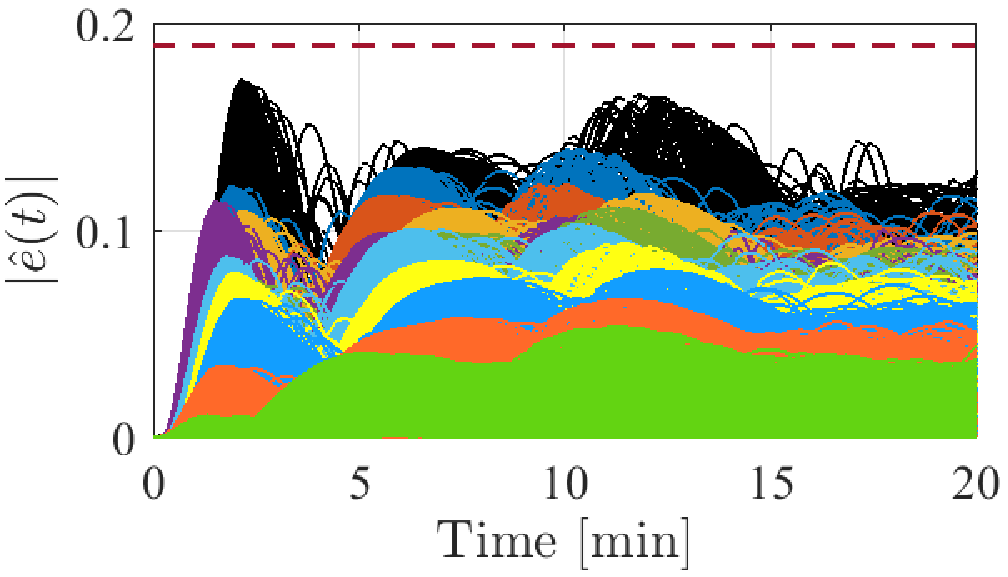}\\
\includegraphics[width=4.5cm]{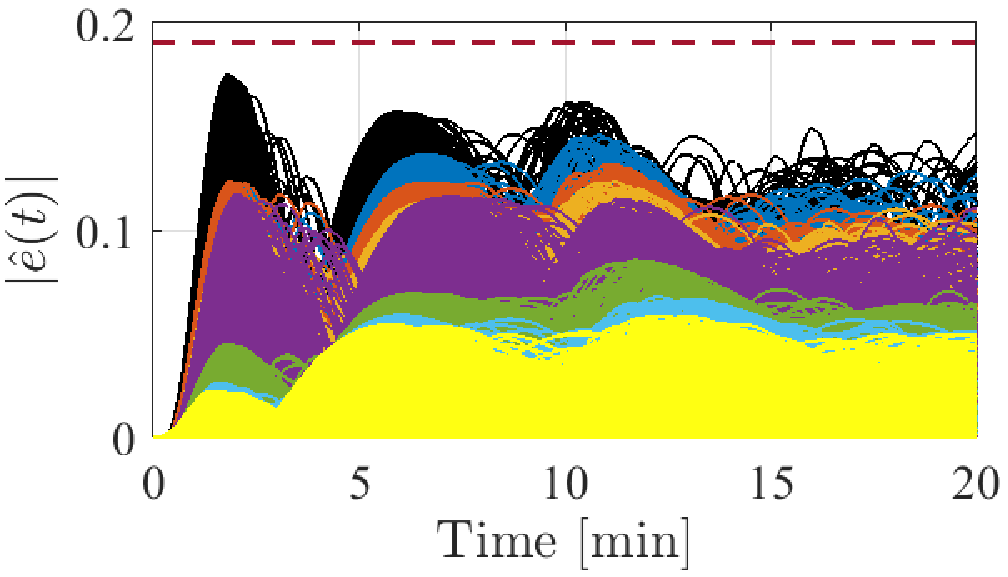}\includegraphics[width=4.5cm]{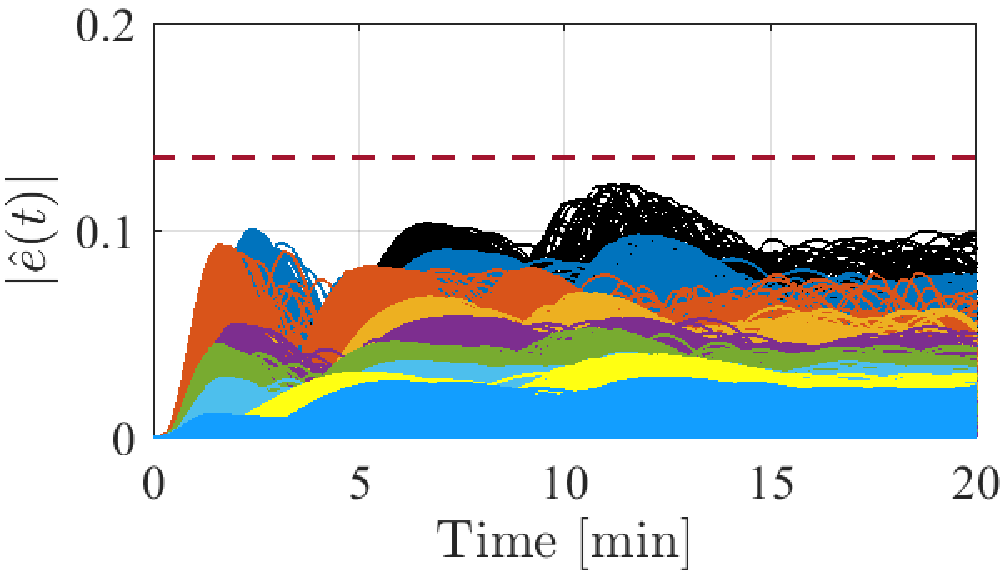}
\caption{Illustration of determining the safety bound $\delta_0$; each color for one patient. Left top: Group 1, right top: Group 2, left bottom: Group 3, right bottom: Group 4.}\label{fig:Illustration0}
\end{figure}

The value of $\delta_0$ can be obtained from extensive simulation studies (Monte Carlo method). \figurename~\ref{fig:Illustration0} shows response of closed-loop anesthesia for 1000 simulations for each patient with successive step-wise reference with random number of steps, levels, and durations (random initial conditions). From this figure, obtained values for $\delta_0$ are reported in \tablename~\ref{Table:approximtion_error}.

Note that although the DSM \eqref{Delta1_1} guarantees constraint satisfaction at all times, it implies that $\tilde{y}(t)\leq0.6-\delta_0,~\forall t\geq0$. Since $0.6-\delta_0<0.5$ for all age groups, by using the DSM \eqref{Delta1_1} there would be no guarantee to reach the desired level of hypnosis, \textit{i.e.}, $v(t)\nrightarrow0.5$.

\begin{figure}[!t]
  \centering
\includegraphics[width=4.5cm]{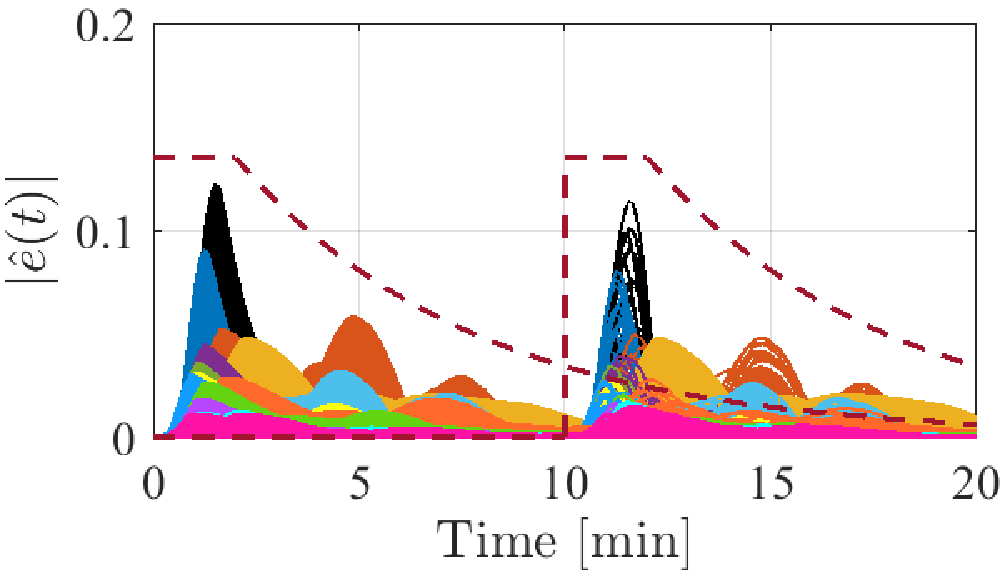}\includegraphics[width=4.5cm]{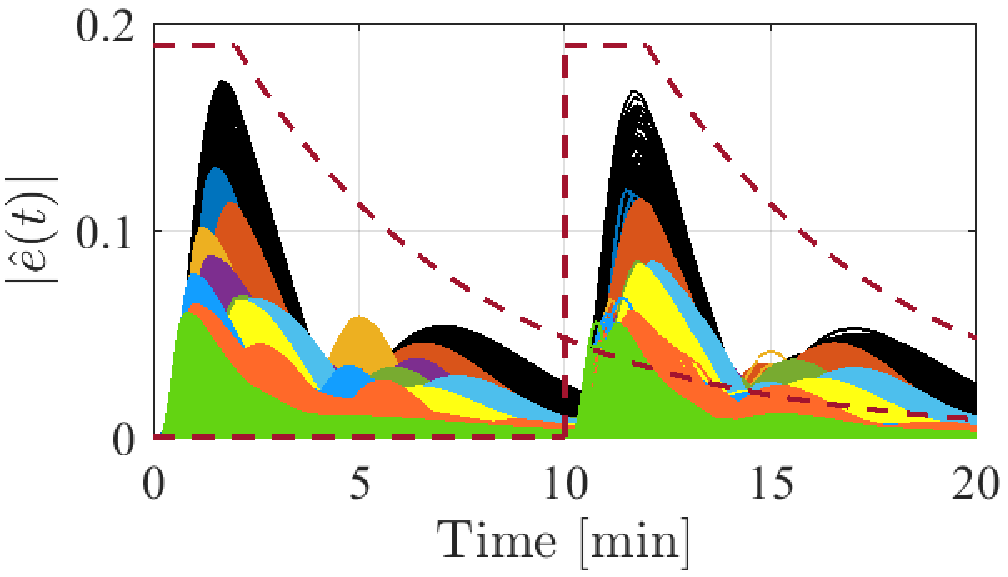}\\
\includegraphics[width=4.5cm]{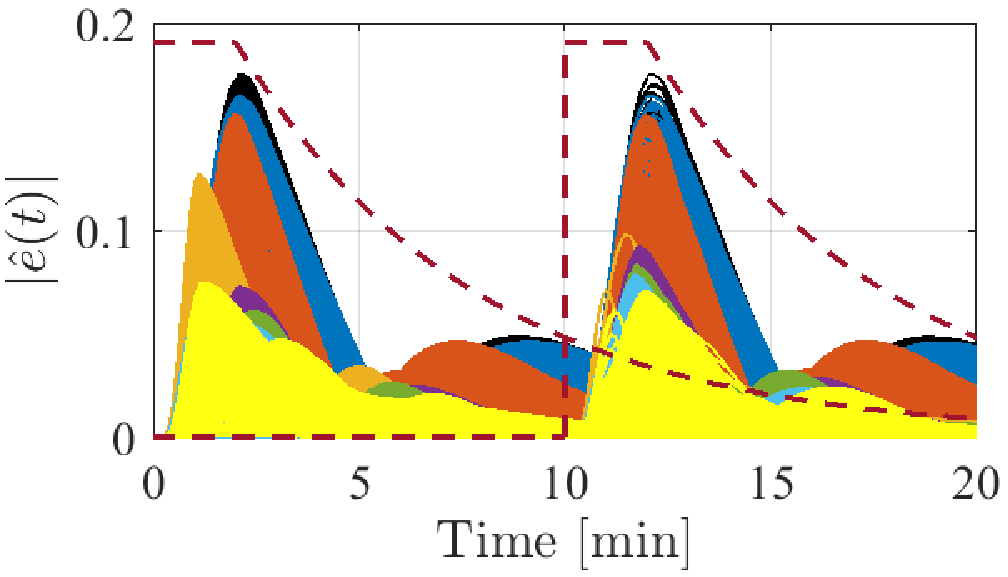}\includegraphics[width=4.5cm]{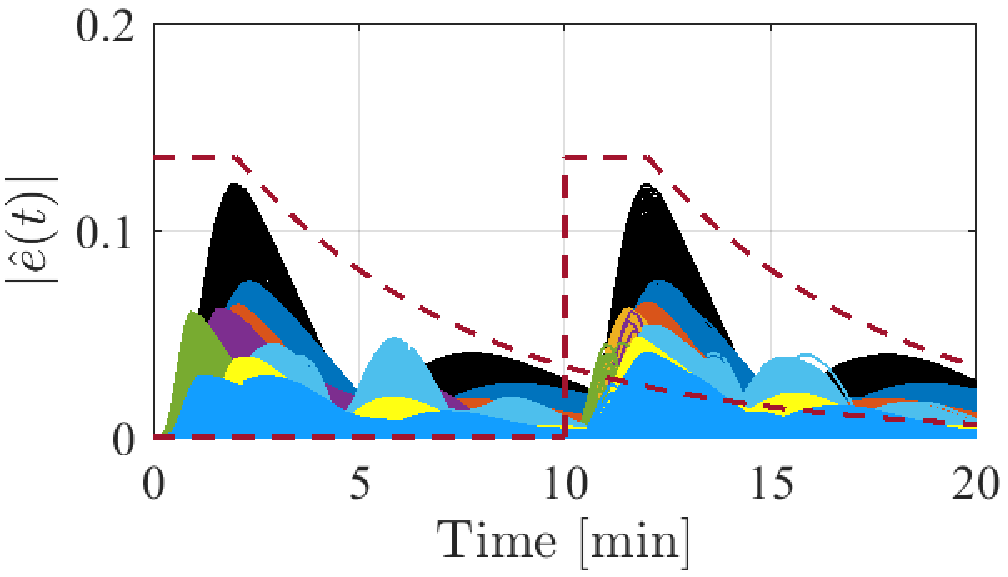}
\caption{Illustration of determining the safety bound $\delta_1$; each color for one patient. Left top: Group 1, right top: Group 2, left bottom: Group 3, right bottom: Group 4.}\label{fig:Illustration1}
\end{figure}

As seen in \figurename~\ref{fig:Illustration1} (1000 simulation studies for each patient with random reference hypnosis level) when $v(t_1)\in[0,0.5]$ (in this figure $t_1=0$ and $10$ [min] to have random initial conditions) and it is kept constant for $t\geq t_1$, we have the following lighter time-varying bound
\begin{equation}\label{eq:propertyError}
|\hat{e}(t)|\leq\delta_1(t),~\forall t\geq t_1,
\end{equation}
where
\begin{align}
\delta_1(t)=&\delta_0\left(u(t-t_1)-u(t-t_1-2)\right)+\delta_0e^{-\frac{t-t_1-2}{5}}u(t-t_1-2),
\end{align}
with $u(t)$ as the step function. Using this bound, the DSM \eqref{Delta1_1} can be replaced by the DSM
\begin{equation}\label{Delta2}
\Delta(t)=\min_{\tau\in\left[t,t+T\right]}\{0.6-\tilde{y}(\tau|t)-\delta_1(t)\},
\end{equation}
where $\tilde{y}(\tau|t)$ can be computed through \eqref{xbar_closedsolution}. Note that the DSM \eqref{Delta2} guarantees reaching the desired level of hypnosis, \textit{i.e.}, $v(t)\rightarrow0.5$ (see \cite{Nicotra2018} for details on the theoretical properties of ERG).

Since the PD model of the patient is not known, to compute $\tilde{y}(\tau|t)$ in \eqref{xbar_closedsolution} we have to use the nominal PD model for each age group. In other words, the matrices $A$ and $B$ in \eqref{Alinear}-\eqref{Blinear} should be computed based on the nominal PD parameters for each age group. Thus, the DSM \eqref{Delta2} can only guarantee constraint satisfaction when applied to the nominal model. In other words, inter-patient variability should be also taken into account in calculating DSM.

\begin{table}[!t]
\centering
\caption{Obtained Values for $\delta_0$.}\label{Table:approximtion_error}
\footnotesize
\begin{tabular}{c|c|c|c|c}
Parameter & Group 1  & Group 2 & Group 3 & Group 4 \\
\hline
$\delta_0$ & 0.1350 & 0.1888 & 0.1907 & 0.1354 \\
\end{tabular}
\end{table}

For this purpose, let define a robustness margin $\delta_2$ as follows
\begin{equation}
\delta_2=\max\limits_{G_1,G_2,G_,G_4}\,\sup\limits_t|\tilde{e}(t)|,
\end{equation}
where $\tilde{e}(t)=\tilde{y}(t)-\tilde{y}_j(t)$ with $\tilde{y}_j(t)$ as the output of the approximated linear system of the \textit{j}-th patient with DSM \eqref{Delta2}. Thus, in order to guarantee constraint satisfaction even in the presence of inter-patient variabilities, it is only needed to further restrict the DSM \eqref{Delta2} with an additional static safety bound as follows
\begin{equation}\label{Delta3}
\Delta(t)=\min_{\tau\in\left[t,t+T\right]}\{0.6-\tilde{y}(\tau|t)-\delta_1(t)-\delta_2\}.
\end{equation}

\begin{figure}[!t]
  \centering
\includegraphics[width=9cm]{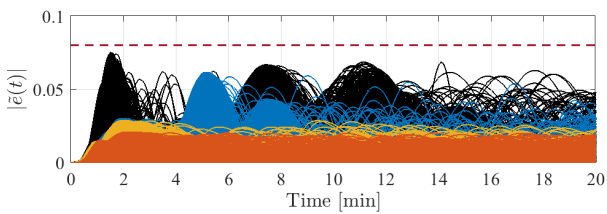}
\caption{Illustration of determining safety bound $\delta_2$; each color for one group.}\label{fig:Illustration2}
\end{figure}

Extensive simulations (1000 simulations for each patient with random reference hypnosis level) have been carried out to determine the value of the safety parameter $\delta_2$. Simulation results for all 44 patients are shown in \figurename~\ref{fig:Illustration2}. As seen in this figure, $\delta_2=0.08$ is sufficient to cover the error caused due to inter-patient variability in all age groups.

It should be remarked that the main advantage of the DSM \eqref{Delta3} is its simplicity and the low computational associated to its computation, which makes the DSM \eqref{Delta3} computable for real-time propofol delivery purposes even on cheap hardware.

In summary, the active set-point prefilter based on ERG framework is implemented as \eqref{ClassicERG} where $\rho(t)$ and $\Delta(t)$ are computed through \eqref{rho} and \eqref{Delta3}, respectively. Note that since the proposed method is designed based on the parameters of 44 patients, in order to make sure that the method is robust against larger uncertainties that might not be seen in simulation studies or may exist in different set of patients, the safety bounds are assumed 5\% bigger than the calculated values.
\section{Results and Discussion}\label{simulation}
Using the proposed ERG solution with $\kappa=10^5$, $\eta=0.01$, all 44 virtual surgeries ($i.e.,$ 44 patients) are simulated in the MATLAB/Simulink environment using the actual PKPD model \eqref{eq:PKmodel}-\eqref{eq:Hill}. For the purpose of comparison, we have also simulated the two-degrees-of-freedom structure without prefilter and with the passive prefilter presented in \cite{Dumont2009}.

Constraint on the amplitude of the propofol infusion rate $I(t)$, is in part due to hard physical constraints on the system. The infusion rate can obviously not be negative, and maximum infusion rate can be enforced to minimize hemodynamic changes. Hence, by assuming that propofol 10 [mg/ml] is being used as the hypnotic drug, the infusion rate is constrained between 0 and 600 [ml/h] \cite{SaraThesis}.

The simulations are run on an Intel(R) Core(TM)i7-7500U CPU 2.70 GHz with 16.00 GB of RAM. The mean computation time for computing $\bar{x}(\tau)$ for $\tau\in[t,T]$ through \eqref{xbar_closedsolution} is 4.768 [ms], which according to the fact that the response time of the propofol delivery system is in the order of minutes, it is largely acceptable for real-time implementation.

\begin{figure}[!t]
  \centering
\includegraphics[width=7.5cm]{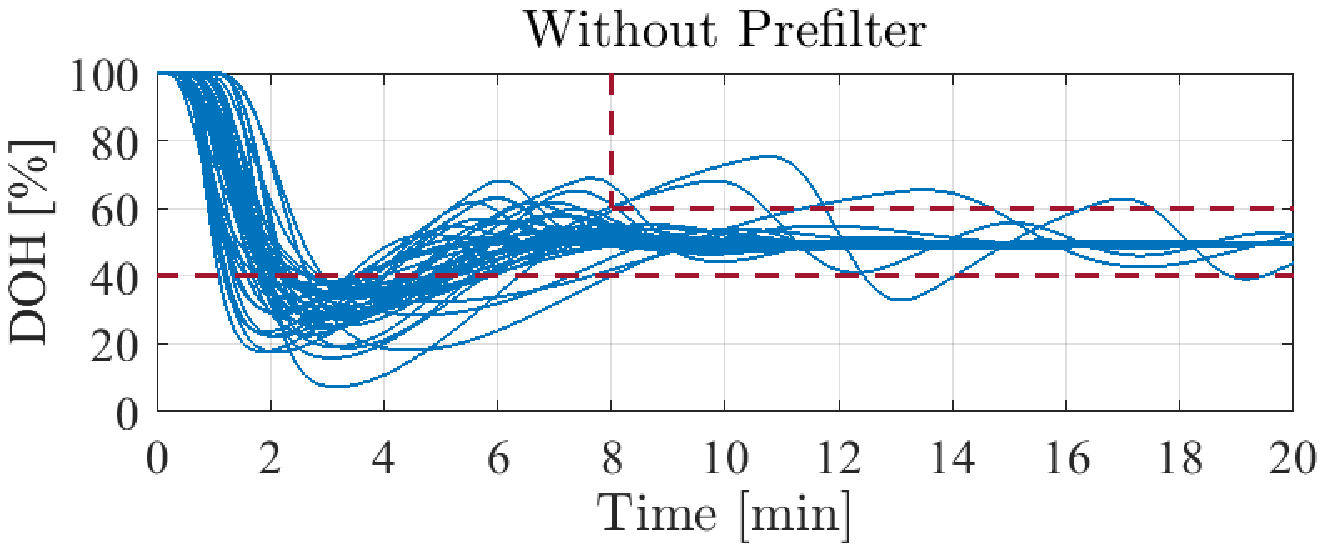}\\
\includegraphics[width=7.5cm]{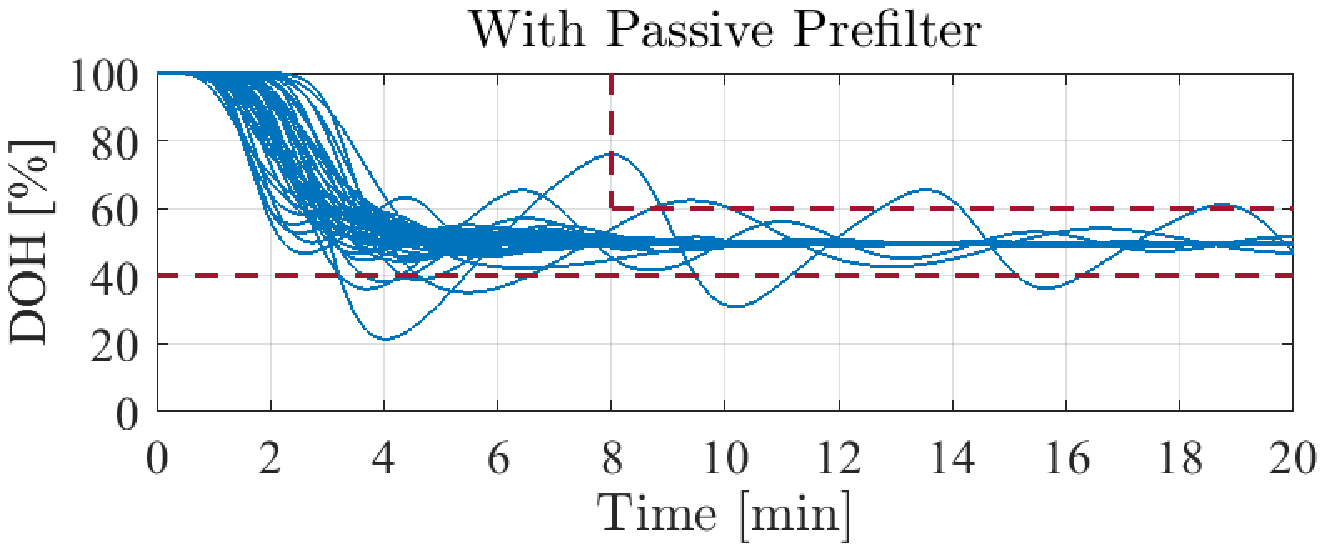}\\
\includegraphics[width=7.5cm]{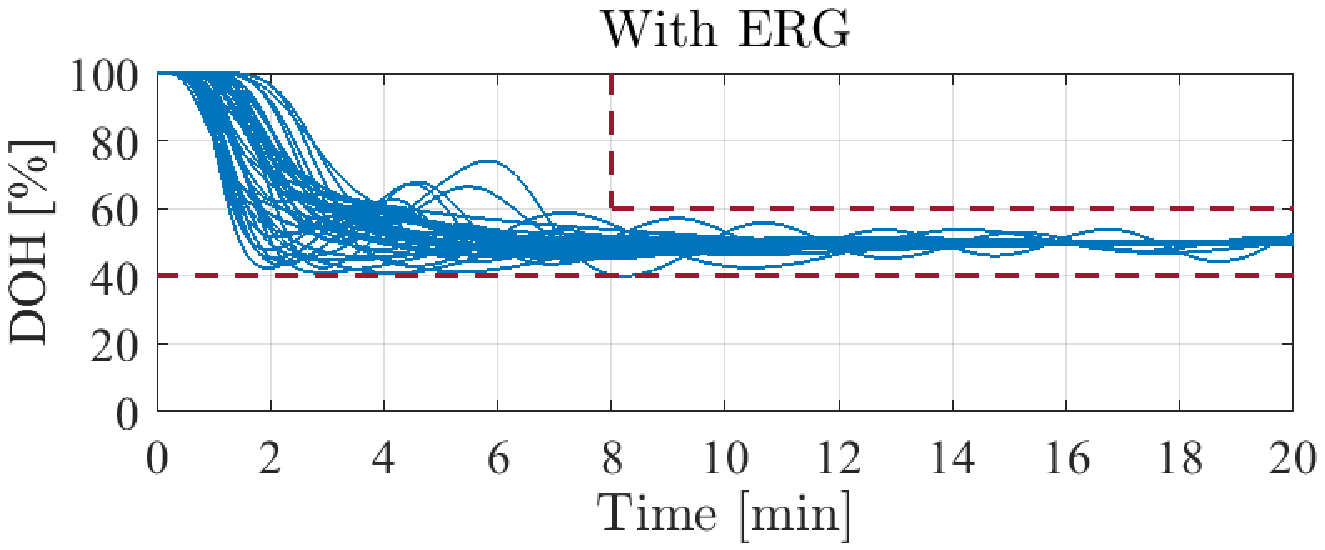}\\
\caption{Closed-loop response of all 44 patients.}\label{Fig:result1}
\end{figure}

\begin{figure}[!t]
  \centering
\includegraphics[width=7.5cm]{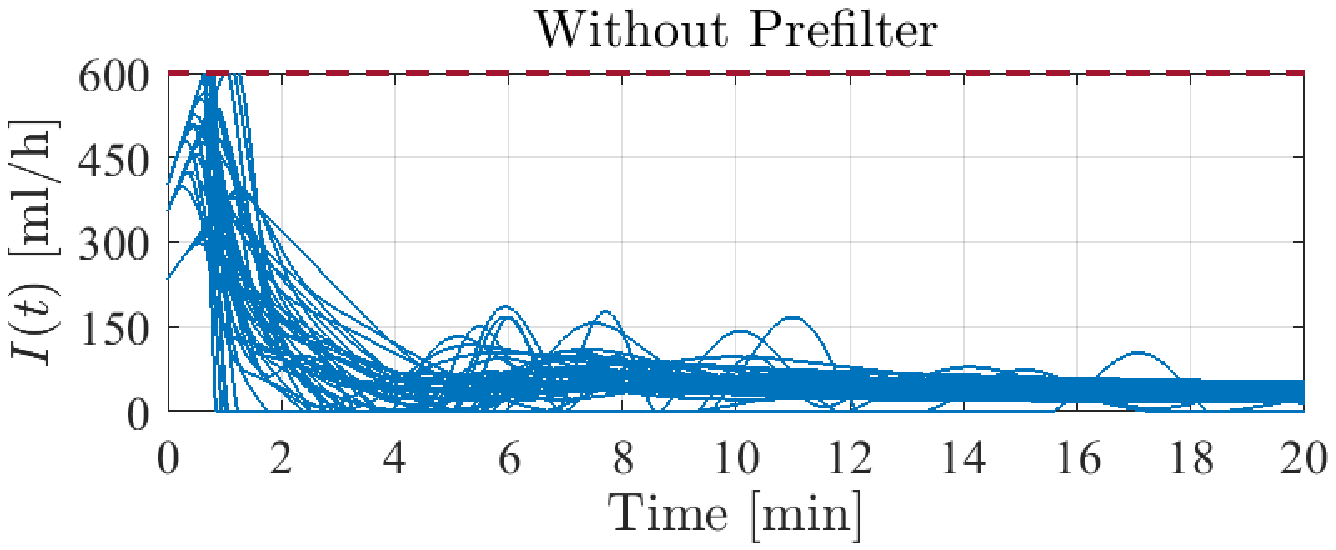}\\
\includegraphics[width=7.5cm]{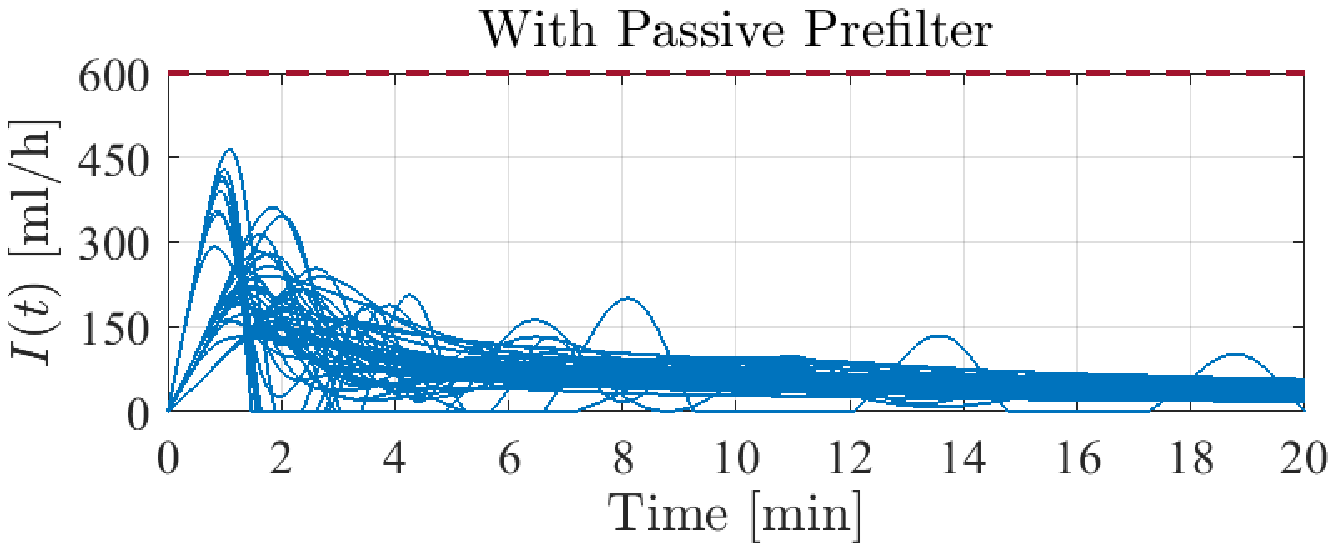}\\
\includegraphics[width=7.5cm]{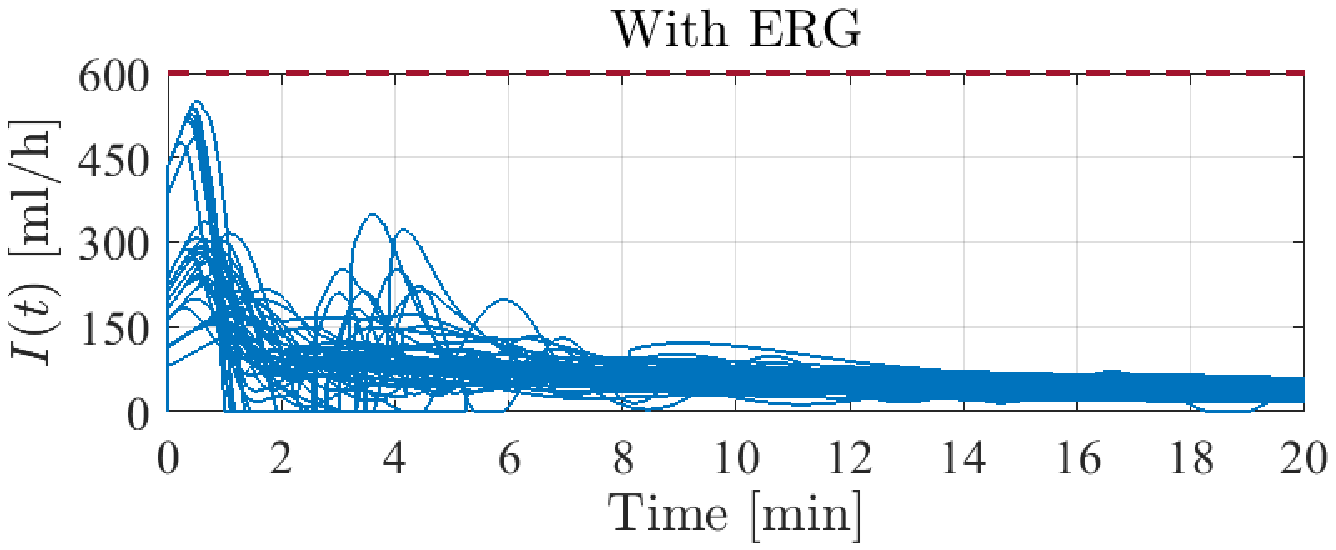}\\
\caption{Infusion rate of all 44 patients.}\label{Fig:result2}
\end{figure}

\begin{figure}[!t]
  \centering
\includegraphics[width=7.5cm]{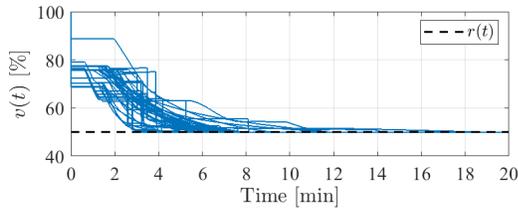}
\caption{Auxiliary reference signal $v(t)$ of all 44 patients.}\label{Fig:result3}
\end{figure}

The resulting closed-loop time responses are shown in \figurename~\ref{Fig:result1}-\ref{Fig:result3}. Note that for the sake of simplicity and convenience, the depth of hypnosis $\text{DOH}(t)$ is defined as follows:
\begin{equation}
\text{DOH}(t)\triangleq100\cdot(1-y(t)),
\end{equation}
where $\text{DOH}(t)=100$ represents a wakeful state and $\text{DOH}(t)=0$ represents the maximum level of hypnosis.

As seen in \figurename~\ref{Fig:result1}-\ref{Fig:result2}, the ERG-based active set-point prefilter can effectively prevent overdosing in all patients while tracking the reference signal. When the passive prefilter is used, five patients in total are in danger of overdosing. Note that without prefiltering the reference signal, all the patients are in the danger of overdosing (see \tablename~\ref{comparison_overdose}).

\begin{table}[!t]
\centering
\caption{Comparison of the Number of Overdosed Patients.}\label{comparison_overdose}
\footnotesize
\begin{tabular}{c|c|c|c}
 & Without  & With Passive & With ERG \\
& Prefilter & Prefilter & \\
\hline
Overdosed Patients & 44 & 5 &  0
\end{tabular}
\end{table}

\begin{table}[!t]
\centering
\caption{Performance Comparison.}\label{comparison1}
\footnotesize
\begin{tabular}{c|c|c|c}
Age Group & Without & With Passive & With ERG \\
& Prefilter &  Prefilter &  \\
\hline
Rise Time \footnotesize{[min]} & & & \\
mean$\pm$SD & 1.71$\pm$0.21 & 5.09$\pm$0.36  & 4.62$\pm$1.24  \\
$[\min,\max]$ & [0.99,2.58] & [2.22,9.80] & [1.47,11.57]\\
\hline
Settling Time \footnotesize{[min]} & & & \\
mean$\pm$SD & 8.53$\pm$1.26 & 6.15$\pm$0.83  & 8.10$\pm$2.08  \\
$[\min,\max]$ & [3.92,35.01] & [3.26,27.68] & [3.83,22.42] \\
\hline
Overshoot \footnotesize{[\%]} & & & \\
mean$\pm$SD & 44.60$\pm$3.59 & 6.27$\pm$3.23  & 9.25$\pm$0.89  \\
$[\min,\max]$ & [24.69,85.42]  & [0.50,57.30] & [2.07,19.99] \\
\end{tabular}
\end{table}

Time-profile of the auxiliary reference signal $v(t)$ when using the proposed technique is presented in \figurename~\ref{Fig:result3}. As seen in \figurename~\ref{Fig:result3}, the ERG acts as an active set-point prefilter that manipulates the auxiliary reference $v(t)$ only when the manipulation does not lead to constraint violation. In simple terms, by using the ERG, instead of applying the desired level of hypnosis instantly, we apply the auxiliary reference $v(t)$ that automatically converges to the desired level of hypnosis so that overdosing prevention is guaranteed at all times.

\tablename~\ref{comparison1} shows the performance of the two-degrees-of-freedom structure without prefilter, with the passive prefilter proposed in \cite{Dumont2009}, and with ERG-based active prefilter, comparing rise time, settling time, and overshoot. SD stands for standard deviation. As seen in this table, settling time and overshoot is increased by using the ERG-based set-point prefilter. However, these increments do not lead to overdosing the patients. More precisely, the active set-point prefilter increases the overshoot and settling time in a wise manner to make sure that the patients will never be in the danger of overdosing.

\begin{table}[!t]
\centering
\caption{Comparison of Used Propofol.}\label{comparison_propofol}
\footnotesize
\begin{tabular}{c|c|c|c}
Age Group & Without & With Passive & With ERG \\
& Prefilter &  Prefilter &  \\
\hline
Used Drug \footnotesize{[ml]} & & & \\
mean$\pm$SD & 30.85$\pm$4.75 & 24.23$\pm$3.86  & 25.76$\pm$4.11  \\
$[\min,\max]$ & [14.99,46.85] & [12.42,35.72] & [14.04,38.56]
\end{tabular}
\end{table}

The amount of used propofol used in first eight minutes to bring the patients to desired level of hypnosis is presented in \tablename~\ref{comparison_propofol}. As seen in this table, the two-degrees-of-freedom structure with ERG-based active set-point prefilter tends to use almost the same amount of propofol of the passive one.

\begin{figure}[!t]
\centering
\includegraphics[width=7.5cm]{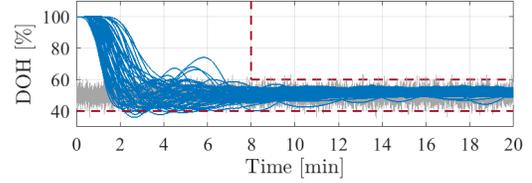}
\caption{Closed-loop response of 44 patients with measurement noise.}\label{fig:DOH_ERG_Noise}
\end{figure}

In order to study the performance of the proposed scheme in the presence of measurement noise, a white noise signal with 0 mean value and 0.1 (20\% of the desired reference) as its variance is added after the Hill saturation element. Closed-loop response is shown in Fig. \ref{fig:DOH_ERG_Noise}. Since to limit the measurement noise a post-processing trending second-order IIR filter was added to the $\text{WAV}_{\text{CNS}}$ monitor \cite{Zikov2006}, measurement noise does not affect the performance of the proposed ERG scheme too much. As seen in Fig. \ref{fig:DOH_ERG_Noise}, although the constraint is violated for some patients, the overall performance is acceptable in the presence of such high measurement noise.

\begin{figure}[!t]
\centering
\includegraphics[width=7.5cm]{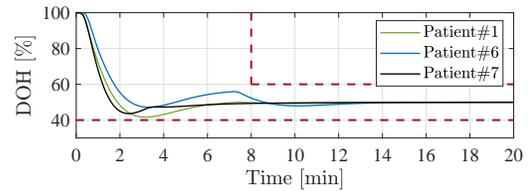}
\caption{Closed-loop response of Patient\#1, \#6, and \#7 given in \cite{Heusden2018}.}\label{fig:DOH_ERG_NEW}
\end{figure}

A new set of data using the NeuroSENSE monitor is identified in \cite{Heusden2018}, which consists of data for 1 patient between 30 and 40 years, 2 patients between 50 and 60 years, and 6 patients above 60 years. Here, we use the proposed method to anesthetize the patients that are in the age range from 18 to 60 years, \textit{i.e.}, Patient\#1, \#6, and \#7. As shown in \figurename~\ref{fig:DOH_ERG_NEW}, the proposed ERG scheme effectively prevents overdosing, while tracking the desired level of hypnosis.

\section{Conclusion}\label{conclusion}
This paper proposed a control scheme structure based on Explicit Reference Governor framework to control the depth of hypnosis during induction phase. Using the proposed scheme, it was show that it is possible to develop an automatic propofol delivery system that can be proven to guarantee overdosing prevention and to provide acceptable performance. The proposed scheme was validated in comparison with a robust PID controller and by simulating 44 patients. The results showed the scheme's effectiveness in controlling the depth of hypnosis and overdosing prevention. More precisely, the proposed scheme exhibited induction phase response with mean rise and settling times of 5 and 8 minutes, as well as mean overshoot of less than 10\%, which meet anesthesiologist's response specifications. Future work includes extending the current work to the development of multi-constraint propofol delivery system in which overdosing and blood pressure changes are addressed as safety constraints, evaluating the proposed constrained control scheme using real clinical trials, and investigating constrained control scheme parameterized in patients' weight. Note that although with a weight-parameterized scheme it is expected to achieve better performance, it requires much more clinical data than those actually available in the literature.

\begin{figure}[!h]
\centering
\includegraphics[width=8cm]{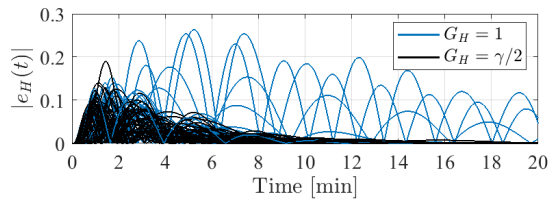}
\caption{Error caused by the Hill linearization as a gain of 1 and of $\gamma/2$.}\label{fig:ErrorH}
\end{figure}

\appendix
The Hill function acts as a time-varying gain. For transient time, the linear gain can be approximated as a straight line passing through the origin and $E_r=0.5$, which gives a constant slope of unity, \textit{i.e.,} $G_H=1$ \cite{Heusden2014}. Around operating regime (\textit{i.e.}, $E_r=0.5$), the linearized gain depends only on the hill steepness coefficient and is equal to $G_H=\gamma/2$. We define the Hill linearization error as $e_H(t)\triangleq y_{nH}(t)-y_{lH}(n)$, where $y_{nH}(t)$ and $y_{lH}(t)$ are the output of the system with nonlinear and linearized Hill function, respectively. \figurename~\ref{fig:ErrorH} shows linearization errors obtained with $G_H=1$ (blue line) and the one obtained with $G_H=\gamma/2$ (black line) for all 44 patients. As seen in this figure, not only linearizing the Hill function around $E_r=0.5$ leads to less error compared with the case with $G_H=1$, but also the error with $G_H=\gamma/2$ converges to zero. The advantage of latter property is revealed in \eqref{eq:propertyError}, where it admits to define a vanishing safety parameter $\delta_1$ to ascertain desired reference tracking.
\balance
\bibliographystyle{IEEEtran}
\bibliography{ref}{}

\end{document}